\documentclass[graybox]{svmult}

\makeatletter
\providecommand*{\toclevel@titlech}{0} 
\edef\toclevel@authorch{\the\numexpr\toclevel@titlech+1} 
\makeatother

\usepackage{type1cm}        
\usepackage{makeidx}         
\usepackage{graphicx}        
\usepackage{multicol}        
\usepackage[bottom]{footmisc}

\usepackage{newtxtext}       %
\usepackage{newtxmath}       

\usepackage{adjustbox}

\usepackage{amsmath}
\usepackage{mathtools}
\usepackage{mathrsfs}

\usepackage[numbers,sort&compress]{natbib}
\usepackage{xcolor}
\usepackage{hyperref}
\hypersetup{
    colorlinks = true,
    allcolors = blue
}

\makeindex             

\begin{document}

\title*{Inverse analysis of asteroseismic data: a review}
\author{Earl P. Bellinger, Sarbani Basu, and Saskia Hekker}
\institute{Earl P. Bellinger \at Stellar Astrophysics Centre, Department of Physics and Astronomy, Aarhus University, Denmark \\
School of Physics, UNSW Sydney, Australia \\
\email{bellinger@phys.au.dk}
\and
Sarbani Basu \at Department of Astronomy, Yale University, USA \\
\email{sarbani.basu@yale.edu}
\and
Saskia Hekker \at Max Planck Institute for Solar System Research, Germany \\
Stellar Astrophysics Centre, Department of Physics and Astronomy, Aarhus University, Denmark \\
\email{hekker@mps.mpg.de}
}
%
%
\maketitle

\abstract*{
Asteroseismology has emerged as the best way to characterize the global and internal properties of nearby stars.
Often, this characterization is achieved by fitting stellar evolution models to asteroseismic observations. 
The star under investigation is then assumed to have the properties of the best-fitting model, such as its age. 
However, the models do not fit the observations perfectly. 
This is due to incorrect or missing physics in stellar evolution calculations, resulting in predicted stellar structures that are discrepant with reality. 
Through an inverse analysis of the asteroseismic data, it is possible to go further than fitting stellar models, and instead infer details about the actual internal structure of the star at some locations in its interior. 
Comparing theoretical and observed stellar structures then enables the determination of the locations where the stellar models have discrepant structure, and illuminates a path for improvements to our understanding of stellar evolution. 
In this invited review, we describe the methods of asteroseismic inversions, and outline the progress that is being made towards measuring the interiors of stars. }

\abstract{
Asteroseismology has emerged as the best way to characterize the global and internal properties of nearby stars.
Often, this characterization is achieved by fitting stellar evolution models to asteroseismic observations. 
The star under investigation is then assumed to have the properties of the best-fitting model, such as its age. 
However, the models do not fit the observations perfectly. 
This is due to incorrect or missing physics in stellar evolution calculations, resulting in predicted stellar structures that are discrepant with reality. 
Through an inverse analysis of the asteroseismic data, it is possible to go further than fitting stellar models, and instead infer details about the actual internal structure of the star at some locations in its interior. 
Comparing theoretical and observed stellar structures then enables the determination of the locations where the stellar models have discrepant structure, and illuminates a path for improvements to our understanding of stellar evolution. 
In this invited review, we describe the methods of asteroseismic inversions, and outline the progress that is being made towards measuring the interiors of stars.
} 

\section{Introduction}
\label{sec:intro}
In stars that are like the Sun, turbulent motions near to the stellar surface randomly excite acoustic modes that oscillate throughout the entire star. 
The oscillation frequencies of these modes depend on the star's internal chemical composition and structure, and particularly on the speeds at which sound travels through the star. 
The sound speed at any given location in the stellar interior is dictated by the local properties of the stellar plasma, and specifically its density, pressure, and compressibility.  
Detection of a sufficient number of these acoustic oscillation modes thus permits inferences of the sound speed---and hence the structure---at various locations within the star. 
As the structure of stars is the principal prediction made by the theory of stellar evolution, asteroseismic measurements of stellar interiors can therefore be used to test the theory. 
This analysis is the topic of the present review.

This is the first review on asteroseismic structure inversions in the post-\emph{Kepler} era \citep[for earlier reviews, see][]{2003Ap&SS.284..153B, 2012AN....333..914C}. 
Here we focus on techniques that have been developed and applied to actual stellar data. 
This review is divided into several, mostly independent sections. 
Section~\ref{sec:inverse} gives a brief mathematical introduction to inverse problems and defines the scope of the review. 
Section~\ref{sec:modelling} describes how to obtain a reference model through an evolutionary inversion, which is generally done before performing a structure inversion. 
Section~\ref{sec:struc} states the structure inversion problem and describes the difficulties and limitations imposed by asteroseismic data. 
Finally, Section~\ref{sec:future} summarizes current results and gives an outlook on future measurements of stellar interiors. 

For a general introduction to asteroseismology, we refer the reader to the two textbooks and numerous review articles on the topic 
\citep[e.g.,][]{2010aste.book.....A, 2017asda.book.....B, 2004SoPh..220..137C, 2007A&ARv..14..217C, 2013AdSpR..52.1581H, 2013ARA&A..51..353C, 2014aste.book...60B, 2017A&ARv..25....1H, 2019arXiv190710457H, 2019LRSP...16....4G, 2019arXiv191212300A}.

\section{A primer on inverse problems}
\label{sec:inverse}
Finish the sequence: 1, 2, 3, 4, \_. 

This task cannot be done with certainty. 
Although the answer would naturally seem to be 5, other solutions are also possible, such as: 
\begin{itemize}
    \item 6 (a sequence known as Ulam's numbers \citep{ulam1962some})
    \item 7 (numbers $n$ such that ${(68 \cdot 10n+7)/3}$ is prime)
    \item 10 (counting in base 5)
    \item 17 ($n$ such that ${n \cdot \varphi(n)}$ is a  palindrome, where $\varphi$ is Euler's totient function)
    \item $-42\pi$ (a sequence we have just invented).
\end{itemize}
Indeed, an infinity of solutions exist, each with infinitely many possible justifications. 

This is often the situation with inverse problems. 
Broadly speaking, inverse problems can be posed for every data-generation scheme, such as those listed above. 
The \emph{forward problem} is to generate data using the scheme (for example, listing the natural numbers). 
The \emph{inverse problem} is to determine the generating scheme itself from the data, or to estimate the parameters of an assumed scheme. 

Inverse problems are characteristically, though not always, \emph{ill-posed}. 
A well-posed problem satisfies three requirements \citep{hadamard}: 
\begin{enumerate}
    \item \textbf{Existence}: a solution exists, 
    \item \textbf{Uniqueness}: there is exactly one solution, and
    \item \textbf{Stability}: the solution changes continuously with changes to the input\footnote{Consider the system of linear equations
\begin{align}
    x &= a + b \label{eq:stab1}\\
    y &= a + b \cdot (1 + \epsilon) \label{eq:stab2}
\end{align}
where $\epsilon$ is an arbitrarily small positive quantity and the rest are real numbers. 
One can imagine that $x,y$ are values observed in nature (i.e., they are the data) and $a,b$ are the hidden parameters that we wish to infer (i.e., they form the data-generation scheme). 
If we observe that $(x,y) = (2,2)$ then clearly $(a,b) = (2,0)$. 
Yet if $(x,y)=(2, 2+\epsilon)$ then $(a,b)=(1,1)$. An arbitrarily small change to the observed value of $y$ has resulted in a completely different solution. The system is unstable.}. 
\end{enumerate}
Ill-posed problems---problems that violate any of these conditions---tend to be difficult or impossible to solve. 
In order to obtain a solution, assumptions must be made, or the problem must be altered. 
An example of a well-posed inverse problem is to find the line corresponding to a few points belonging to that line. 
An ill-posed inverse problem would be to find a line after any of the points have been perturbed by noise. 
As no line will pass through all of the points, an assumption must be made, such as assuming the noise is Gaussian-distributed and accepting the least-squares solution.

For more information, the reader is referred to textbooks and review articles on inverse problems \citep[e.g.,][]{kirsch2011introduction, neto2012introduction, tenorio2001statistical}.

\subsection{Inverse problems in asteroseismology} 
There are numerous inverse problems in asteroseismology.  
These include inversions to infer a star's: 
\begin{enumerate}
    \item \textbf{Evolutionary history}: age, mass, initial composition, and evolutionary parameters such as mixing efficiencies, assuming the theory of stellar evolution is approximately correct \citep[e.g.,][]{pulone1997age, 2016ApJ...830...31B, 2017EPJWC.16005003B, 2019A&A...622A.130B, 2020MNRAS.491.4752B, 2017ApJ...839..116A, AngelouAccepted, 2019PASP..131j8001H, 1994ApJ...427.1013B, 2004ApJ...600..419G, 2009ApJ...699..373M, 2015MNRAS.452.2127S, 2017ApJ...835..173S, 2019MNRAS.484..771R, 2019ApJ...887L...1B, 2019MNRAS.490.2890J, 2018ApJS..237...15A, 2019MNRAS.486.4612B, 2020MNRAS.492L..50B}.
    \item \textbf{Internal structure}: specifically, the isothermal sound speed at various locations around the stellar core \citep{2017ApJ...851...80B, 2019ApJ...885..143B}.
    \item \textbf{Radial differential rotation}: rates of rotation of the stellar core and surface \citep[e.g.,][]{2012ApJ...756...19D, 2014A&A...564A..27D, 2015A&A...580A..96D, 2015MNRAS.452.2654B, 2019A&A...631L...6E, AhlbornAccepted}. 
    \item \textbf{Latitudinal differential rotation}: surface rotation rates at different latitudes \citep[][]{2018Sci...361.1231B, 2018A&A...619L...9B, 2019A&A...623A.125B}. 
    \item \textbf{Integrated quantities}: any quantity that takes its value over some or all of the stellar interior \citep{1998MNRAS.297L..76P}, such as the star's 
        \begin{enumerate}
            \item \textit{Mean density}: $(3/4\pi) M/R^3$, where $M$ is the mass of the star and $R$ is its radius \citep{2012A&A...539A..63R}. 
            \item \textit{Acoustic radius}: $\int c^{-1}\,\text{d}r$, where $c=(\Gamma_1 P/\rho)^{1/2}$ is the speed of sound, $\Gamma_1$ is the adiabatic compressibility, $P$ is pressure, $\rho$ is density, and $r$ is the distance to the stellar center \citep{2015A&A...574A..42B}. 
            \item \textit{Sound speed gradient}: $\int r^{-1} (\text{d}c/\text{d}r)\, \text{d}r$, which is a proxy for stellar age on the main sequence because the sound speed gradient depends on the mean molecular weight, which changes over time due to hydrogen fusion \citep{2015A&A...583A..62B, 2016A&A...596A..73B}. 
        \end{enumerate}
\end{enumerate}
These problems are ill-posed and therefore difficult to solve---some due to existence, some due to uniqueness, some due to stability, and some due to a combination of these. 
This review focuses on the first two of these problems, and especially the second. 
That being said, essentially all of these problems can be solved using the techniques discussed in this review.

\section{Evolutionary inversions}
\label{sec:modelling}
The first step for most inversions is to obtain a reference model. 
This is in fact itself an inverse problem. 
Given observations of a star, we seek to determine its age, internal structure, and other parameters using the theory of stellar evolution. 

Stellar evolution theory can be regarded as the function
$f_{\text{evol}}(\tau, \mathbf x) = \mathbf y$, 
where $\tau$ is the age of the star and $\mathbf x = [$mass, chemical composition, etc.$]$ are the initial conditions. 
Other parameters may include anything relevant to the evolution, such as the mixing length and overshooting parameters for convection, diffusion coefficients, nuclear reaction rates, opacities, stellar companions, and so on. 
The forward problem is to compute a stellar model\footnote{There are numerous software implementations for computing $f_{\text{evol}}$; one open-source and freely available one-dimensional stellar evolution code is \textsc{Mesa} \citep{2011ApJS..192....3P}. 
The global oscillation modes of the resulting stellar model is generally computed using a separate code. 
Two freely available codes for computing linear adiabatic stellar oscillation frequencies are \textsc{Adipls} \citep{2008Ap&SS.316..113C} and \textsc{Gyre} \citep{2013MNRAS.435.3406T}. } that has observables $\mathbf y = [$effective temperature, metallicity, oscillation frequencies, etc.$]$. 
Other observables may include the radius, luminosity, and so on.  

We seek to invert this function: given observations $\mathbf y$, we wish to infer $f_{\text{evol}}^{-1}(\mathbf y) = [\tau, \mathbf x]$. 
However, this inverse function is not guaranteed to exist because $f_{\text{evol}}$ is not necessarily injective. 
Broadly speaking, there are three approaches for solving the evolution inversion problem:
\begin{enumerate}
    \item \textbf{Machine learning}: an approximation to the inverse function $f_{\text{evol}}^{-1}$ is learned from a large number of examples of $\tau, \mathbf x$, and $\mathbf y$ \citep{pulone1997age, 2016ApJ...830...31B, 2017EPJWC.16005003B, 2019A&A...622A.130B, 2020MNRAS.491.4752B, 2017ApJ...839..116A, AngelouAccepted, 2019PASP..131j8001H}. 
    \item \textbf{Optimization}: sometimes referred to as repeated forward modeling, a likelihood function that quantifies how well a given model fits the data is maximized, often along with priors on the parameters of the model \citep[e.g.,][]{1994ApJ...427.1013B, 2004ApJ...600..419G, 2009ApJ...699..373M, 2015MNRAS.452.2127S, 2017ApJ...835..173S, 2019MNRAS.484..771R, 2019ApJ...887L...1B, 2019MNRAS.490.2890J, 2018ApJS..237...15A}. 
    \item \textbf{Scaling relations}: by assuming homology with the Sun \citep[e.g.,][]{2019arXiv190710457H}, the age and mass of a star can be deduced using calibrated scaling relations \citep{2019MNRAS.486.4612B, 2020MNRAS.492L..50B}. 
    The chemical composition can be approximated using a galactic chemical evolution law \citep[e.g.,][]{2017ApJ...835..173S}. These parameters can then be used with $f_{\text{evol}}$ to obtain a reference model. 
\end{enumerate}
In past decades, when observations consisted of only the effective temperature, metallicity and sometimes luminosity, this problem was ill-posed because the solution lacked uniqueness, i.e., many stellar models satisfied the same constraints. 
With the advent of asteroseismology, this problem is ill-posed because the solution lacks existence: there are no stellar models which fit all the asteroseismic data \citep[][]{2017ApJ...851...80B, 2019ApJ...885..143B}. 
This is the case even for the Sun and models of solar evolution \citep[e.g.,][]{2016LRSP...13....2B, 2019ApJ...881..103Z}.

This motivates the structure inversion problem. 
If no stellar model has the correct structure, then what is the correct structure?

\section{Sound speed inversions} \label{sec:struc}

The problem at hand is to deduce the internal structure of a star, insofar as it is possible, from asteroseismic measurements, without assuming the theory of stellar evolution. 
This problem is generally cast as one of deducing the differences in the internal sound speed profile between the star and a suitably chosen reference model, which can be achieved through an analysis of the differences in their frequencies of oscillation. 
This approach leads directly to an assessment of the quality of stellar evolution models, and particularly whether the models have the correct structure at the locations within the star where the inverse analysis is successful. 
This problem is not fundamentally different from the structure inversion problem in helioseismology \citep[e.g.,][]{1991sia..book..519G, 2016LRSP...13....2B} and geoseismology \citep{1967GeoJ...13..247B, 1968geoj...16..169b}. 
However, inverting asteroseismic data comes with unique difficulties and limitations over helioseismic inversions \citep{2003Ap&SS.284..153B, 2017asda.book.....B}. 

The fact that the solar disk can be resolved permits the detection of solar oscillation modes with spherical degrees of $\ell \lessapprox 250$ \citep[][]{2016LRSP...13....2B}. 
In contrast, cancellation effects limit asteroseismic detections to modes with $\ell \leq 3$ \citep[e.g.,][]{2010aste.book.....A, 1977AcA....27..203D, 2015MNRAS.452.2127S, 2017ApJ...835..172L}. 
Although all acoustic oscillation modes have upper turning points near to the stellar surface, only the high degree modes have lower turning points in the outer parts of the star \citep[see Figure 4.4 of][]{2017ApJ...851...80B}. 
Thus, unlike the Sun \citep[e.g.,][]{2009ApJ...699.1403B}, only the near-core region ($\sim 0.05-0.35R$ away from the center) can be probed with asteroseismic inversions.

The structure inversion problem is complicated by the fact that frequencies of stellar oscillations do not depend exclusively on the sound speed; the effects of a second variable, such as the density or adiabatic compressibility, must be considered as well \citep[e.g.,][Section 3.3.3.2]{2010aste.book.....A}. 
Unfortunately, asteroseismic mode sets are too limited to resolve (or suppress) the effects of a second variable \citep[][Chapter~10]{2017asda.book.....B}. 
A path forward was realized very early \citep{1993ASPC...40..541G}. 
By assuming an equation of state, it is possible to explain oscillation frequencies in terms of the abundance of helium and any function of pressure and density, such as the squared isothermal sound speed $u=P/\rho$. 
The reason this is helpful is because the helium abundance only substantially affects the oscillation frequencies in ionization zones, which are located very near to the stellar surface where asteroseismic inversions are not sensitive anyway. 
Hence, asteroseismic data probes the $u$ profile of the stellar core.

A further difficulty concerns the fundamental parameters of the star. 
The solar fundamental parameters, and particularly the solar mass and radius, are precise, accurate, and known independently of helioseismology. 
For stars, these parameters must be estimated, generally also from the same asteroseismic data with which one wishes to perform the inverse analysis. 
The relative uncertainties of asteroseismic mass and radius estimates are about 4\% and 2\%, respectively \citep[e.g.,][]{2019A&A...622A.130B}. 
Thus, the model we choose as reference may not have the same mass and radius as the star. 
This has two consequences. 
The first is in the mode frequencies themselves: the frequencies scale with the stellar root mean density, and so a difference there results in a constant relative difference between the frequencies of the star and of the stellar model. 
Secondly, a difference in mass or radius affects the $u$ profile itself. 
From a dimensional analysis of the equations of stellar structure, one finds ${P \propto M^2 / R^4}$ and ${\rho \propto M/R^3}$, and hence ${u \propto M / R}$. 
Thus, asteroseismic structure inversions are actually sensitive to the dimensionless structure $u' = u R/M$ \citep[][Figures 9 and 10]{2003Ap&SS.284..153B}. 
Hereinafter, $u'$ will be referred to simply as the sound speed.

A final complication is that theoretical and observed mode frequencies are well-known to systematically differ due to deficiencies in modeling the near-surface layers of stars \citep[e.g.,][]{2017EPJWC.16002001B}. 
Generally, these differences are merely fit and subtracted. 
Though clearly not the best solution, it does not otherwise present a major hurdle here. 

Once a suitable reference model is selected, one can perturb the equations of adiabatic stellar oscillations and linearize them around the reference model \citep[e.g.,][]{2003Ap&SS.284..153B, 2010aste.book.....A, 1991sia..book..519G, 2011LNP...832....3K, 2016LRSP...13....2B, 2017asda.book.....B, buldgen2017development, 2018PhDT.......125B, 2018ASSP...49...75R}. From this, one obtains the central equations of the asteroseismic structure inversion---a set of equations, one for each observed oscillation mode, that express how the mode frequencies depend on the stellar structure and composition: 
\begin{equation} \label{eq:inversion}
    \color{darkgray}\underbrace{\color{black}\frac{\delta\nu_i}{\nu_i}}_{\mathrlap{\color{darkgray}\text{difference in oscillation frequency}}}\color{black}
    = 
    \int \color{darkgray}\overbrace{\color{black}
    K_i^{(u',Y)}\, 
    \frac{\delta u'}{u'} 
    +
    K_i^{(Y,u')}\, 
    \delta Y}^{\mathclap{\color{darkgray}\text{differences in structure \& composition}}}\color{black}
    \,
    \text{d}r
    +
    \color{darkgray}\underbrace{\color{black}F_{\text{surf}}(\nu_i)}_{\mathclap{\color{darkgray}\text{surface term}}}\color{black}
    +
    \color{darkgray}\overbrace{\color{black}\epsilon_i,}^{\mathclap{\color{darkgray}\text{measurement errors}}}\color{black}
    \qquad 
    i=1\ldots N. 
\end{equation}
Here the index $i$ refers to a mode whose frequency $\nu_i$ has been measured. 
The difference in any quantity between the star and the reference model is denoted by $\delta$. 
The kernels $K_i$ for a particular mode are functions derived from the perturbation analysis that relate $u'$ and the fractional helium abundance $Y$ to that mode's frequency \citep[for examples, see][Figures 1.21 and 1.22]{2018PhDT.......125B}. 
The surface term $F_{\text{surf}}$ is some function of frequency \citep[e.g.,][]{2014A&A...568A.123B}. 
The unknown measurement error, given by $\epsilon_i$, is assumed to be drawn from normal distribution with zero mean and standard deviation $\sigma_i$. 

The principal assumption of a structure inversion is that the mode kernels of the reference model serve as sufficient approximations to the actual mode kernels of the star. 
As mode kernels do not differ much between even rather different models \citep{2000ApJ...529.1084B}, the inversion results can be considered independent of evolutionary considerations, despite usually using a model calculated using stellar evolution theory as reference. 

The problem now at hand is to determine the $\delta u'/u'$ (or, equivalently, $u'$) profile from the data using these $N$ equations. 
This problem is ill-posed for two reasons. 
The first is uniqueness: infinitely many sound-speed profiles fit the data equally well \citep[e.g.,][]{1991sia..book..519G}. 
These include nonphysical solutions, such as those with negative sound speeds. 
The second is stability: as the modes trace similar paths through the star, the kernels are nearly linearly dependent, and so a small perturbation to the mode frequencies can lead to a disproportionate change in the result (recall Equations~\ref{eq:stab1} and \ref{eq:stab2}). 
Thus the problem must be changed. 
A way forward was discovered by geologists Backus and Gilbert \citep{1967GeoJ...13..247B, 1968geoj...16..169b}: the method of \emph{optimally localized averages} (OLA).

\subsection{The method of optimally localized averages} 
\label{sec:OLA}

OLA is a method that can be used to measure the sound speed at various locations within a star. 
The method works by combining the mode kernels belonging to the observed modes into a new kernel, called the \emph{averaging kernel}, which by design is only sensitive to one region of the star, called the \emph{target radius}. 
If the creation of an averaging kernel is successful, it can be used as an instrument to measure the sound speed at the target radius. 
However, the creation of an averaging kernel must be balanced against the amplification of the data errors and the contamination of the effects of a differing helium abundance. 
It is important to note that the OLA method does not involve fitting the data; in fact, the construction of the averaging kernel does not make use of the oscillation frequencies at all; it uses only their uncertainties.

The averaging kernel $\mathscr{K}$ is formed by taking a linear combination $\mathbf c$ of the mode kernels: $\mathscr{K} = \sum_i^N c_i K_i^{(u', Y)}$. 
If this linear combination is chosen such that $\mathscr{K}$ integrates to unity and only has amplitude around the target radius, then $\int \mathscr{K} \delta u'/u' \, \text{d}r$ provides a localized average, denoted $\langle \delta u' / u' \rangle$, of the sound speed at the target radius. 
Now consider a sum over all the equations in Equation~\ref{eq:inversion}: 
\begin{equation} \label{eq:OLA}
    \color{darkgray}\underbrace{\color{black}\left\langle \frac{\delta u'}{u'} \right\rangle}_{\mathrlap{\color{darkgray}\text{localized average}}}\color{black}
    \simeq
    \sum_i^N \color{darkgray}\overbrace{\color{black}c_i \frac{\delta\nu_i}{\nu_i}}^{\mathclap{\text{observed data}}}\color{black}
    =
    \int \color{darkgray}\underbrace{\color{black}\mathscr{K}  \, \frac{\delta u'}{u'}}_{\mathclap{\text{averaging term}}}\color{black} \, \text{d}r 
    +\int \color{darkgray}\overbrace{\color{black}\mathscr{C}  \, \delta Y}^{\mathclap{\text{cross term}}}\color{black} \, \text{d}r 
 + \sum_i^N \color{darkgray}\underbrace{\color{black}c_i \, F_{\text{surf}}(\nu_i)}_{\mathclap{\color{darkgray}\text{surface effects}}}\color{black}
 + \sum_i^N \color{darkgray}\overbrace{\color{black}c_i \, \epsilon_i.}^{\mathclap{\text{errors}}}\color{black}
\end{equation}
The same linear combination that forms the averaging kernel is applied to all terms, such as the cross-term kernel $\mathscr{C} = \sum_i c_i K_i^{(Y, u')}$. Thus, $\mathbf c$ must be chosen in such a way that suppresses these other effects.

\begin{figure}[t]
    \centering
    \makebox[0.5\textwidth][c]{%
        \adjustbox{trim={0cm 8mm 0cm 0cm},clip}{%
            \includegraphics[width=0.5\textwidth]{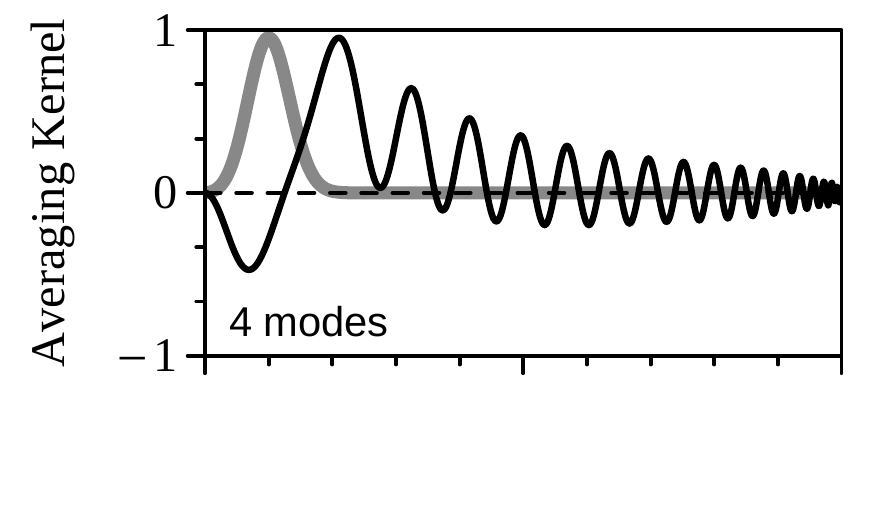}%
        }%
        \adjustbox{trim={1.1cm 8mm 0cm 0cm},clip}{%
            \includegraphics[width=0.5\textwidth]{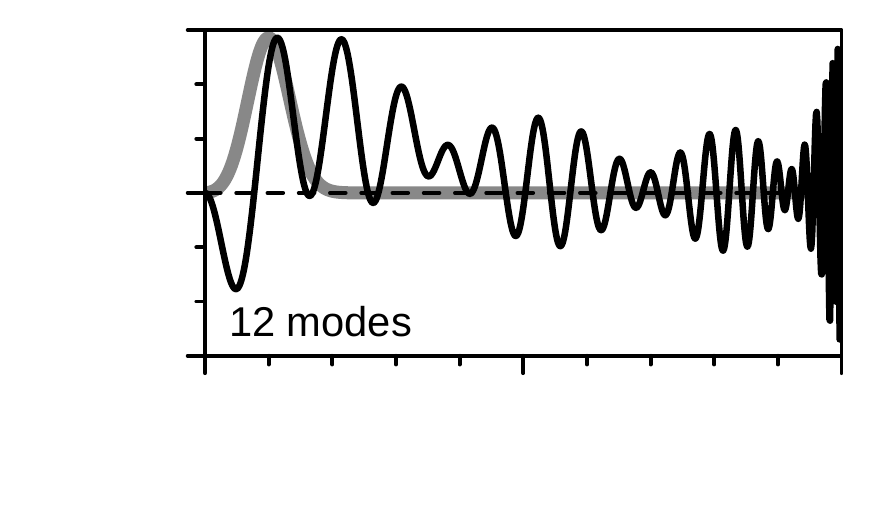}%
        }%
    }\\%
    \makebox[0.5\textwidth][c]{%
        \adjustbox{trim={0cm 0mm 0cm 0cm},clip}{%
            \includegraphics[width=0.5\textwidth]{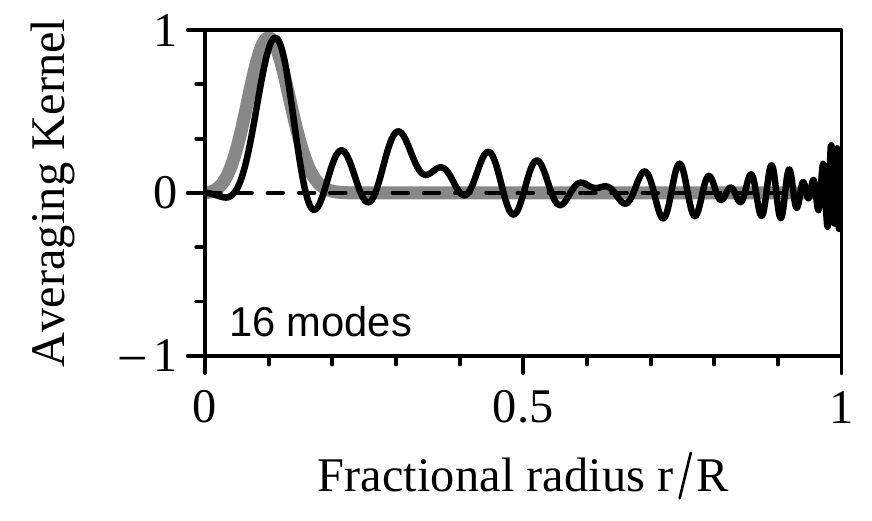}%
        }%
        \adjustbox{trim={1.1cm 0mm 0cm 0cm},clip}{%
            \includegraphics[width=0.5\textwidth]{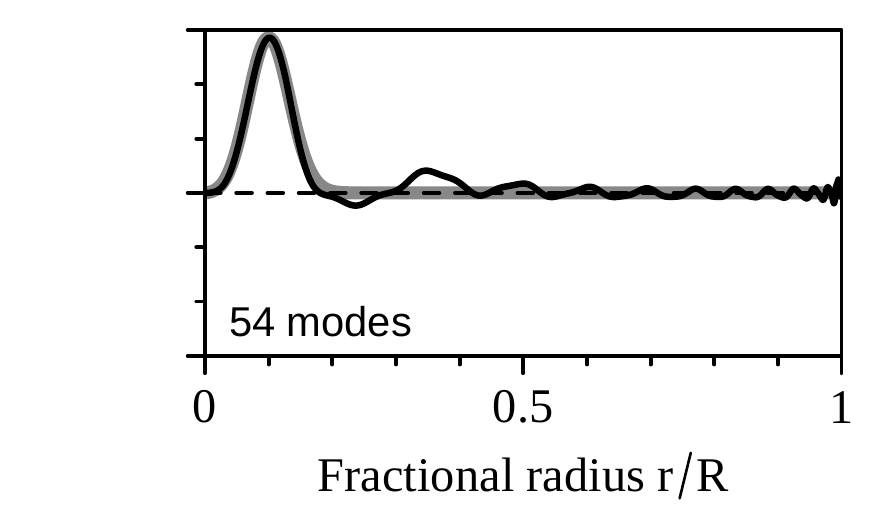}%
        }%
    }\\%
    \caption{Averaging kernels at a target radius of $0.1R$ constructed using four differently sized mode sets from the solar-like oscillator 16~Cyg~A \citep{2017ApJ...851...80B}. 
    The target kernel is indicated by the gray line. }
    \label{fig:avg_kerns}
\end{figure}

A common way of solving the OLA problem is known as Subtractive OLA \citep[SOLA,][]{1994a&a...281..231p}. 
This method defines a \emph{target kernel}, which is typically a Gaussian function peaked at the target radius. 
Free parameters control the width of the target kernel, the suppression of the cross-term kernel, and the amplification of uncertainty \citep[e.g.,][]{1999MNRAS.309...35R}. 
SOLA then finds the $\mathbf c$ that makes the averaging kernel resemble the target kernel, insofar as it is possible \citep[e.g.,][]{2018PhDT.......125B}. 
Examples of averaging kernels constructed using SOLA with \emph{Kepler} data \citep{2017ApJ...835..172L} are given in Figure~\ref{fig:avg_kerns}. 
The final averaging kernel is reasonably well-localized and suffices to estimate $\delta u'/u'$ at the target radius.

\section{Summary of current results and future outlook} \label{sec:future}
Prior to the widespread detection of oscillations in solar-like stars by space missions such as \emph{Kepler}, a number of early works investigated the prospect of asteroseismic inversions using faux data generated from stellar models \citep{1993ASPC...40..541G, 1998mons.proc...33G, 2001ESASP.464..407B, 2001ESASP.464..411B, 2002ESASP.485..337R, 2002ESASP.485..249B, 2002ESASP.485...95T, 2002ESASP.485...75R}. 
These works are summarized in the first review on asteroseismic inversions \citep{2003Ap&SS.284..153B}. 
Generally, these early attempts were optimistic about the number of mode frequencies that would be observed and the uncertainties on those observations. 
Most critically, the assumed uncertainties on the mass and radius of the target star were generally too small. 

To date, asteroseismic structure inversions have been applied to three actual stars: the solar analogs 16~Cyg~A and B \citep{2017ApJ...851...80B} and a star with a small convective core, KIC~6225718 \citep{2019ApJ...885..143B}. 
In the former case, the dimensional $u$ profile was sought by inverting many reference models spanning the uncertainties in mass and radius. 
A procedure called inversions-for-agreement was devised, which had the benefit of using the data to select the parameters of the inversion. 

In each of the three cases, there was tension between the predicted and observed internal stellar structures. 
The model of the close solar analog 16~Cyg~B performed the best, while the star with a convective core---the least solar-like star---showed the greatest differences. 
For this star, alternate models employing differing physics in the evolution of the star were calculated. 
The input physics of convective core overshoot and atomic diffusion were tested against the asteroseismic observations. 
However, no set of input physics was found that could explain the observed structure. 

The field of asteroseismic inversions is still in its infancy with many interesting results yet to come. 
Several more stars from legacy \emph{Kepler} data are still amenable to inverse analysis; this is work in preparation from the present authors. 
By analyzing a large number of stars, it will be possible to rigorously test theories of stellar evolution across a variety of non-solar regimes and pinpoint the areas of stellar modeling that are currently deficient. 
Future space missions will furthermore yield orders-of-magnitude more detections of solar-like oscillations, at which point inverse analyses will likely need to be applied at a large scale rather than star-by-star. 

As a final note, the stars discussed here are all firmly on the main sequence. Structure inversions of evolved solar-like stars, whose non-radial acoustic modes couple with gravity modes in the core and give insights into the very deep interiors of stars, are sure to be a promising source of discovery.

\begin{acknowledgement}
These proceedings are dedicated to the memory of Michael J.\ Thompson. 
Funding for the Stellar Astrophysics Centre is provided by The Danish National Research Foundation (Grant agreement no.: DNRF106). 
S.H. acknowledges funding from the European Research Council under the European Community’s Seventh Framework Programme (FP7/2007-2013) / ERC grant agreement no 338251 (StellarAges). S.B. acknowledges partial support from NSF grant AST-1514676 and NASA grant NNX13AE70G. 
\end{acknowledgement}

\bibliographystyle{spphys}
\bibliography{Bellinger}

\end{document}